\documentstyle[epsfig,psfig,12pt]{article}
\newcommand{\beq}{\begin{equation}}
\newcommand{\eeq}{\end{equation}}
\newcommand{\bea}{\begin{eqnarray}}
\newcommand{\eea}{\end{eqnarray}}
\title{Simulation of nuclear effects in quasi elastic and resonant
neutrino interactions}
\author{}
\date{\empty}

\begin{document}
\maketitle
\begin{center}
G.Battistoni$^1$, P.Lipari$^2$, J.Ranft$^3$, E.Scapparone$^4$\\
{\em 1. INFN, Sezione di Milano, I--20133 Milano (Italy)}\\
{\em 2. INFN, Sezione di Roma and Dipartimento di Fisica dell' Universit\' a
``La Sapienza'', I--00185 Roma (Italy)} \\
{\em 3. Universit\"at Siegen, Fachbereich Physik, D--57068 Siegen, 
         Germany}\\
{\em 4. INFN, Lab. Nazionali del Gran Sasso, I--67010 Assegi (AQ), Italy}
\end{center}
\vskip 3. cm
\begin{abstract}
The effects of nuclear re-interactions in quasi elastic and resonant
neutrino interactions have been considered in the framework of
the nuclear models of the DPMJET code.
A preliminary investigation on the modifications induced on the final state
has been performed. Some consequences affecting the experimental
identification are discussed.
\end{abstract}
\vskip 3cm
\newpage
\section{Introduction}

The interest in neutrino interaction is nowadays 
mostly oriented to the question of neutrino mass and flavour oscillations. 
The anomaly in atmospheric neutrino flux, pointed out by 
many experiments\cite{kamio,imb,soudan}, suggested the possibility to explore 
the region $\Delta m^{2}$$\simeq$ $10^{-2}$-$10^{-3}$ e$V^{2}$,
using a Long Base Line (LBL) neutrino beam. Considering 
the recent CHOOZ result\cite{chooz}, ruling out the $\nu_{\mu}$$\rightarrow$
$\nu_{e}$ explanation for the atmospheric anomaly, 
the interest for LBL
experiments is shifted to $\nu_{\tau}$ appearance search 
with energy $E_{\nu}$$\simeq$4--10 GeV, since the recent result from 
Superkamiokande\cite{superk}, folded with the older 
Kamiokande and IMB results, suggests oscillation
parameters $sin^{2}2\theta$$\simeq$1 and  $\Delta m^{2}$$\simeq$
5$\cdot$$10^{-3}$. The $\nu_{\tau}$ tagging is rather difficult at these
energies both from a statistical and an experimental point of view.
The first point comes from the overlap of the 
low cross section at these energies with
the poor neutrino flux in the LBL case, with respect to a Short Base Line
situation, decreasing like 
$\simeq$ $r^{-2}$, 
where r is the distance
between the neutrino production point and the neutrino detection point.
The second point is related to the intrinsic difficulties in measuring
the kinematic variables of
$\nu_{\tau}$ interaction products at low energy.

However, at these neutrino energies, the selection of quasi elastic and
resonant events can be useful
to identify $\nu_\tau$ candidates. At energies not far from the threshold for
the production of $\tau$ lepton, 
 the quasi elastic cross section gives still a large contribution to 
the the total charged current cross section. 
 Moreover the intrinsic simplicity of the kinematics, simplifies
the search ``a la Nomad''\cite{nomad} for $\nu_{\tau}$ identification. 
Favourable
are those events in which the $\tau$ lepton decays into 
an electron  or muon (plus neutrinos) or into a single pion (plus neutrino). 

However, the intrinsic cleanness offered by kinematics of 
the quasi--elastic interaction of neutrinos on single 
nucleons, can be obscured in case of nuclear targets.
Nuclear effects, like Fermi motion and nuclear 
re--interaction of the nucleon inside the nucleus, 
have to be considered with care.
These effects can provide momentum imbalance, 
the production of additional particles and may lead to 
the misidentification of the outgoing nucleon. 
All this phenomenology can deteriorate the signal/background ratio,
the event recognition efficiency  and  the vertex identification.
Moreover, in case of experiments that plan to use 
emulsions  as  active  targets,  one  expects to  see, sometimes,
low  energy  particles  and  fragments close 
to  the  interaction vertex  (``grey'' and ``black" particles, in 
the language  of emulsions).                                

Guided by this consideration we 
start from the  quasi--elastic and resonant neutrino  
scattering simulation code
of \cite{lipari}, which already takes into account some nuclear  effects,
like Fermi motion and Pauli blocking. We have interfaced it to 
the 
DPMJET--II--.4.1 code \cite{DPMJETII,Ranftsare95,Dpmjet23}, 
which  was   created to  
treat hadron--nucleus and  nucleus--nucleus interaction  
at high energy  
but which also contains a Formation Zone Intranuclear Cascade
(FZIC) for low energy interaction of the produced secondaries
with the spectator nucleons, nuclear evaporation and the
formation of the residual nucleus. 
The nuclear models of
DPMJET have  been  tested against a  wide set  of  experimental data. In
particular,  the  main  check of  the  de--excitation  and  fragmentation
algorithms  comes from  the  comparison with  emulsion 
data \cite{Ferrari95a,Ferrari96a}. 


More  refined models  which take into
account the effect of the nuclear potential on the particle trajectories
exist\cite{fluka3,fluka1,fluka2,fluka4,fluka5} 
and will be considered by other authors\cite{FerrariIca}
also 
in the framework
of neutrino simulation. While waiting for these more
refined codes, we think that we can obtain from the use of DPMJET a lot
of essential information, which, as a first approximation, is reasonable
and not achievable from presently existing codes.

In this paper we present the first 
results obtained using this code, on the 
comparison with single nucleon interaction with and without
considering the nuclear effects and the implication for 
$\nu_{\tau}$ explicit search.

\section{Description of the code}
The code extends the quasi--elastic 
neutrino--nucleon and neutrino--nucleus
model QEL\cite{lipari},
based on the formulation of
Lewellyn--Smith\cite{smith},
to neutrino collisions on nuclear targets including the 
intranuclear cascade and nuclear fragmentation and evaporation.
The QEL code has been  transformed to double precision,
to match the DPMJET
environment,  and it has been modified to use the 
Fermi momenta of the nucleons
from DPMJET.

All neutrino flavours can be considered. 
In the case of tau lepton production, the leptonic
decay into muon (or electron) plus neutrinos can be optionally activated.
For this reason, we have also introduced the calculation of the
polarization of the final lepton, according to the treatment
of \cite{albright}.

The nucleon generated by  the QEL code can re--interact in DPMJET
inside
the nucleus according to the Formation Zone Intranuclear Cascade (FZIC)
model \cite{Stodolski75,Ranft89a,Ferrari95a} 
contained inside DPMJET.
Secondaries from this first collision 
are followed along straight trajectories and
may induce in turn
intranuclear cascade processes if they reach the end of their
formation zone inside
the target,
otherwise they leave the nucleus without interaction.
Inelastic secondary interactions in the FZIC are
described by the Monte Carlo code HADRIN \cite{HADRIN,Haenssgen84a}.
For the sampling of elastic nucleon--nucleon 
scattering below $4~GeV$ the
parameterization of the HETC--KFA code 
\cite{HETC-KFA,HETC-KFA1} was adopted.

The treatment of nuclear effects within the MC model has
already been discussed in~\cite{Moehring91} and in more detail
in \cite{Ferrari95a,Ferrari96a}. 
Since they are essential in
calculating excitation energies of nuclei left after primary
interactions and intranuclear cascade processes we summarize the 
basic ideas.
Fermi momenta for nucleons as well as a simplified treatment of the
nuclear potential are applied to control the
generation of low--energy particles. Nucleon momenta are sampled from
zero--temperature Fermi distributions
\begin{equation}
\frac{dN^{\mbox{\scriptsize n,p}}}{dp}=N^{\mbox{\scriptsize n,p}}
\frac{3p^2}{(p_{\mbox{\scriptsize F}}^{\mbox{\scriptsize n,p}})^3}.
\end{equation}
The maximum allowed Fermi momenta of neutrons and protons are
\begin{equation}
\label{maxFermi}
p_{\mbox{\scriptsize F}}^{\mbox{\scriptsize n,p}}=
\left[\left(\frac{N^{\mbox{\scriptsize n,p}}}{V_A}\right)
\frac{3h^3}{8\pi}\right]^{\frac{1}{3}}
\end{equation}
with $V_A$ being the volume of the corresponding nucleus with an
approximate nuclear radius $R_A=r_0A^{1/3}, r_0=1.29$~fm.

Modifications of the actual nucleon momentum distribution, as they would
arise, for instance, 
taking the reduced density and momenta in the nuclear 
skin into consideration,
effectively result in a
reduction of the Fermi momenta as compared to those sampled from
Eq.~(\ref{maxFermi}). This effect can be estimated by a correction
factor $\alpha_{\mbox{\scriptsize mod}}^{\mbox{\scriptsize F}}$ which 
modifies the Fermi--momenta. 
Results presented in this paper have been obtained with 
$\alpha_{\mbox{\scriptsize mod}}^{\mbox{\scriptsize F}}$=0.60.
The depth of the nuclear potential is assumed 
to be the Fermi energy and the
binding energy for outer shell nucleons
\begin{equation}
\label{nucpot}
V^{\mbox{\scriptsize n,p}}=
\frac{(p_{\mbox{\scriptsize F}}^{\mbox{\scriptsize n,p}})^2}
{2m_{\mbox{\scriptsize n,p}}}+
E^{\mbox{\scriptsize n,p}}_{\mbox{\scriptsize bind}}.
\end{equation}
To extend the applicability of the model to the energy region well below
1~GeV an approximate treatment of the Coulomb--potential is provided.
The Coulomb--barrier modifying the nuclear potential is calculated from
\begin{equation}
\label{coulpot}
V_{\mbox{\scriptsize C}} = \frac{e^2}{4\pi\epsilon_0 r_0}
\frac{Z_1Z_2}{(A_1^{1/3}+A_2^{1/3})}
\end{equation}
with the mass numbers $A_1, A_2$ and charges $Z_1, Z_2$ of the
colliding nuclei, i.e. with $A_1=|Z_1|=1$ for charged hadrons entering
or leaving the target nucleus. $e$ denotes the elementary charge and
$r_0=1.29$~fm.

The excitation energy $U$ of the residual nucleus with mass number 
$A_{\mbox{\scriptsize res}}$
and charge $Z_{\mbox{\scriptsize res}}$, i.e. the energy above the 
ground state mass $E_{0,\mbox{\scriptsize res}}$, is given as
\begin{eqnarray}
&U&=E_{\mbox{\scriptsize res}}-E_{0,\mbox{\scriptsize res}},\nonumber\\
&E_{0,\mbox{\scriptsize res}}&=
Z_{\mbox{\scriptsize res}}m_{\mbox{\scriptsize p}}+
(A_{\mbox{\scriptsize res}}-Z_{\mbox{\scriptsize res}})m_{\mbox{\scriptsize n}}
-E_{\mbox{\scriptsize bind}}
(A_{\mbox{\scriptsize res}},Z_{\mbox{\scriptsize res}}).
\end{eqnarray}
The binding energy 
$E_{\mbox{\scriptsize bind}}
(A_{\mbox{\scriptsize res}},Z_{\mbox{\scriptsize res}})$ is obtained
using the experimentally determined excess 
masses of all known (measured) 
nuclides and using mass formulae for
nuclides far from the stable region, 
where no measurements are available.
The excitation energy is obtained within our model from an explicit 
consideration of the effects of
the nuclear potential (Eq.~(\ref{nucpot})) and the 
Coulomb energy (Eq.~(\ref{coulpot})), i.e. from corrections 
which are applied to the
4--momenta of the final state hadrons leaving the spectator nucleus. 
We modify the energies of these hadrons by the potential barrier and
rescale the 3--momenta correspondingly. It is assumed
that these corrections have to be applied to nucleons wounded in
primary and secondary interactions and only to those hadrons which are
formed inside the spectator nucleus corresponding to the sampled
formation time. Among these particles we find apart from the nucleons a
small fraction of other baryons, which are assumed to move in a nucleon
potential and mesons to which we apply an effective meson potential
of 0.002~GeV. Due to energy--momentum conservation these
corrections lead to a recoil momentum and, therefore,
to an excitation of the residual nucleus. In addition, there is
a further contribution to the recoil momentum of the residual nucleus
arising from potential corrections applied to the momentum of the
projectile hadron entering the nuclear potential and from cascade
nucleons with kinetic energies below the nuclear potential which are
therefore not able to escape the spectator nucleus.
Pauli blocking is considered, and events are rejected accordingly.
Instead, no angular momentum barriers are considered.

At the end of the intranuclear cascade 
the residual nucleus is supposed to
be left in an equilibrium state, 
in which the excitation energy $U$ is shared
by a large number of nucleons. Such an equilibrated prefragment 
nucleus is 
supposed to be characterized by its mass, charge, and excitation energy 
with no further memory of the steps which led to its formation.
The excitation energy can be higher than the separation energy,
thus nucleons and light fragments ($\alpha$,d,$^3$H, $^3$He) can
still be emitted: they constitute the low--energy (and most
abundant) part of the emitted particles in the rest system of the
residual nucleus, having an average energy of few MeV.

In heavy nuclei the evaporative process is in 
competition with another equilibrium
process, that is fission~\cite{Vandenbosh73}.
For the fission probability,
a statistical method can be used~\cite{Weisskopf37,Bohr39}.

Other de--excitation mechanisms are more suitable 
for light residual nuclei. The one adopted for this calculations
is the so called Fermi Break--up model~\cite{Fermi50,Epherre67},
 where the excited nucleus is supposed to
disassemble just in one step into two or more fragments, with branching
given by plain phase space considerations. 
According to the picture of the compound nucleus like an equilibrated
system determined only by its mass,
charge and excitation energy, with no
memory of previous steps of the interaction, Fermi Break--up is activated
in the model every time the current compound nucleus has mass number
$A\le 17$, including possible light fission fragments.
The fragmentation of higher mass compound nuclei is not yet included
in the model. This process, although its cross section is quite
small, is important when considering the distribution of residual
nuclei, because it can produce isotopes very far both from the target
mass and from the fission product distribution.

The evaporation stage ends when the nuclear 
excitation energy becomes lower 
than all separation energies for nucleons and fragments. This residual 
excitation energy is then dissipated through emission of photons. 
$\gamma$--de--excitation proceeds through a cascade of 
consecutive photon emissions, until the ground state is reached. The 
cascade is assumed to be statistical as long as the excitation energy 
is high enough to allow the definition of a continuous nuclear level 
density. Below a (somewhat arbitrary) threshold, set at the pairing gap 
value, the cascade goes through transitions between discrete levels. 
In reality, photon emission occurs even during the preequilibrium and 
evaporation stages, in competition with particle emission, but its
relative probability is low, and it is presently neglected in the model.
\section{Simulation results}
In order to test the features of the code, 
we considered 5,000 quasi elastic interactions
of 10 GeV $\nu_{\mu}$ on light (Carbon), intermediate (Silicon)
 and heavy (Iron) targets.
\begin{table}[hbt]
\begin{center}
\begin{tabular}{|l|c|c|c|c|c|c|}
\hline
&&&&&&\\
Target & $<N_{n,p}>$ & $<N_{\pi^{\pm}}>$& $<N_{\gamma}>$ 
& $<$ $E_{\gamma}$ $>$ &P(n=2) & $<$ $P_{\perp}^{miss}$ $>$  \\
 & & & & (MeV) & & (MeV/c) \\
\hline
free nucleons & 1. & 0 &0&0      & 100$\%$       &   0      \\
\hline
Carbon & 1.49 $\pm$0.02 &0.016 $\pm$0.002 &0.79 $\pm$0.02 & 2.1   &27.4$\%$ 
&   175       \\
\hline
Silicon & 2.02 $\pm$0.02 &0.025 $\pm$0.002 & 2.68 $\pm$0.02  &1.5
& 0.$\%$        &   183      \\
\hline
Iron   & 2.26 $\pm$0.05 &0.023 $\pm$0.005 & 3.42 $\pm$0.06 & 
1.5  &     0. $\%$  &   216     \\
\hline
\end{tabular}
\caption{\it Comparison of the result of $\nu_{\mu}$ interaction 
($E_{\nu_\mu}$=10 GeV) on different targets.}
\end{center}            
\end{table}
We immediately see that additional particles appear in the final state.
Table 1 shows the comparison of the results 
on different targets, as far as
the average nucleon, gamma and charged pion multiplicities 
and the average
de--excitation gamma energy are concerned. Moreover, we show  
the probability $P$ to obtain a ``clean'' 
two body final state (just muon
plus proton) and the average missing $P_{\perp}$.
We notice that the main effect of the nuclear re--interaction consists in
additional nucleon production, although at very low momentum, as it
will be shown later. Charged pion production(whose kinetic energy is
around 300 MeV) is rather negligible.
\begin{figure}[htcb]
\centerline{\epsfig{file=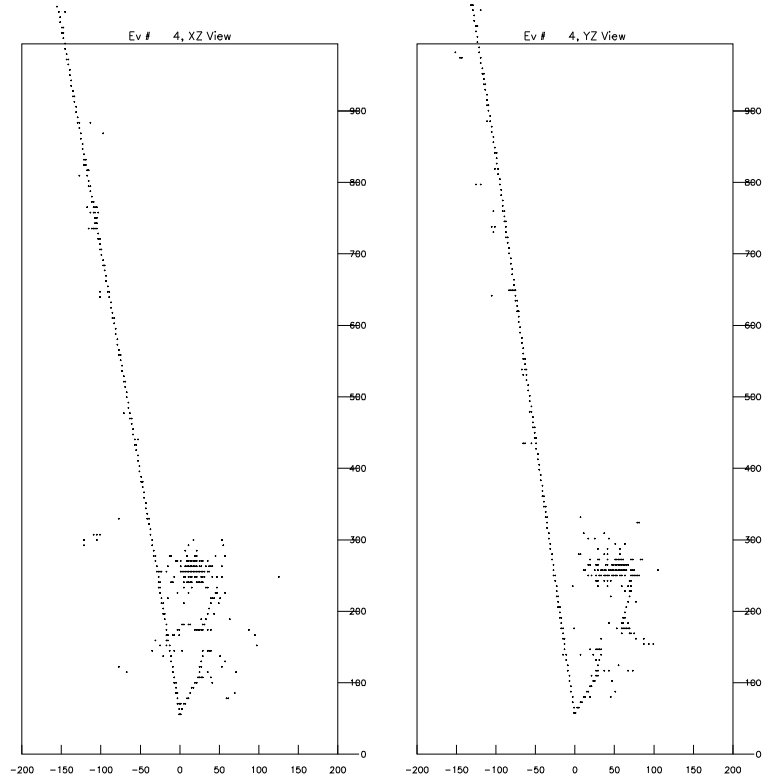,width=10cm}}
\caption { Example of a simulated quasi elastic $\nu_\mu$ event at 10 GeV
  in a glass--scintillator sampling calorimeter (nuclear
  re--interaction off). } 
\label{f:fig1}
\end{figure}     
\clearpage
To give an idea of the event topology distortion introduced by nuclear 
effects, we simulated a fine grain low density sampling
calorimeter (1/4 X$_0$ glass  + 1.5 cm of liquid scintillator) using the
FLUKA package\cite{fluka2}.
Fig. 1 and 2 show the same QEL event, without nuclear effects (Fig. 1) and
and with nuclear effects (Fig. 2).                                       
\begin{figure}[htbc]
\centerline{\epsfig{file=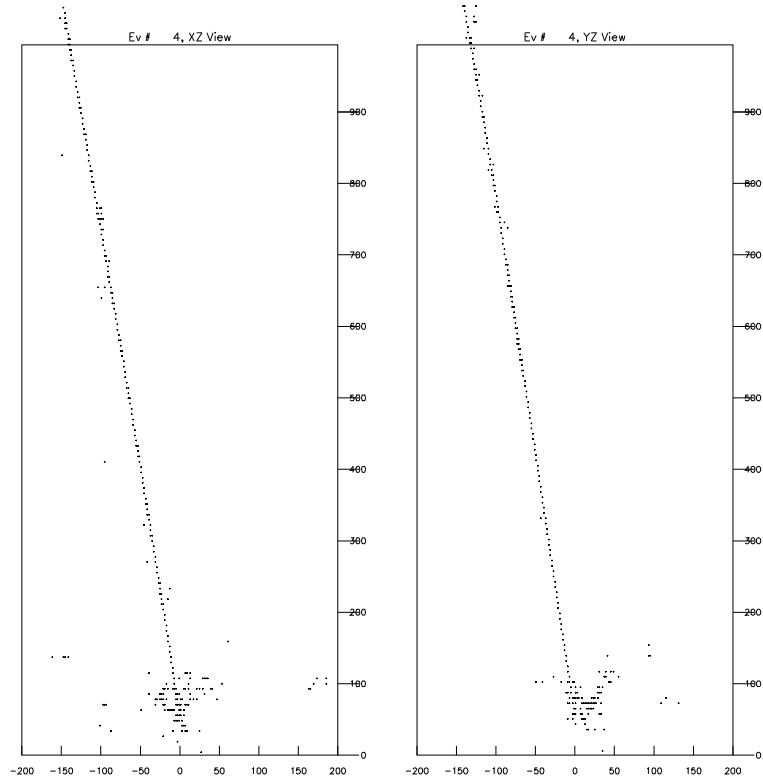,width=10cm}}
\caption { The same event of Fig. \protect\ref{f:fig1} when nuclear
re--interactions are considered.}
\label{f:fig2}
\end{figure}       
\begin{figure}[htb]
\centerline{\epsfig{file=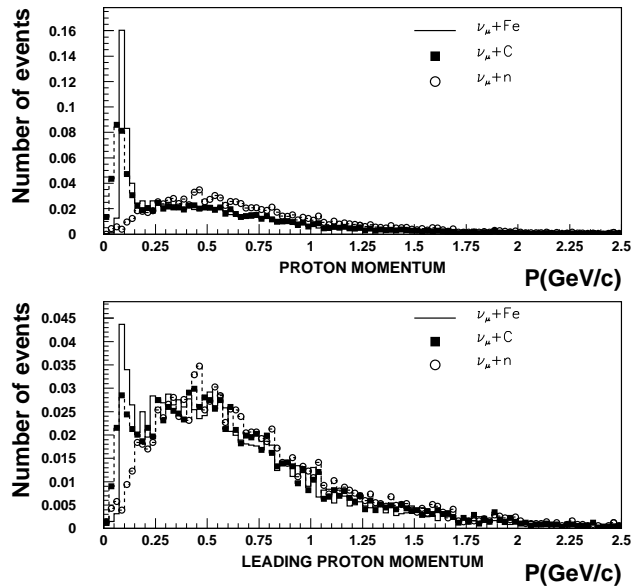,width=9cm}}
\caption {Protons (top) and leading proton(bottom)
momentum distribution, in 
$\nu_{\mu}$+n, $\nu_{\mu}$+C and 
$\nu_{\mu}$+Fe interaction ($E_{\nu_\mu}$ = 10 GeV)}
\label{f:fig3}
\end{figure}
It's evident, looking at these figures, that 
the identification and tracking of the proton becomes
more difficult, both for the presence of extra particle 
in the final state and for the modification of the kinematics of
the proton itself.

Fig. 3 shows the momentum of the protons produced in the
interaction. In the top part we consider all the protons emitted while
in the bottom part, we considered only the event leading proton
(the most energetic one). We show in the same figure the case with and
without nuclear re--interaction. We learn that the bulk of the additional
protons is produced with low momentum, as typical of 
intranuclear cascade products.
As expected, the effect tends to become more significant for heavy target
nuclei.  
There is also some dependence on energy.
The effects of nuclear re--interaction manifest themselves in the distortion
of the kinematic variables. 
As an example, in Fig. 4 we show the distribution of the angle between the muon
and the outgoing leading proton. The case with and without nuclear 
re--interactions are distinguished and compared to the case of the 
free neutron target. We notice how the proton now,
even the leading one,
can access much larger angle with respect to the simple kinematics of
pure neutrino--neutron quasi--elastic scattering. Further kinematic
effect can be observed looking at event missing $P_{\perp}$.
\begin{figure}[htb]
\centerline{\epsfig{file=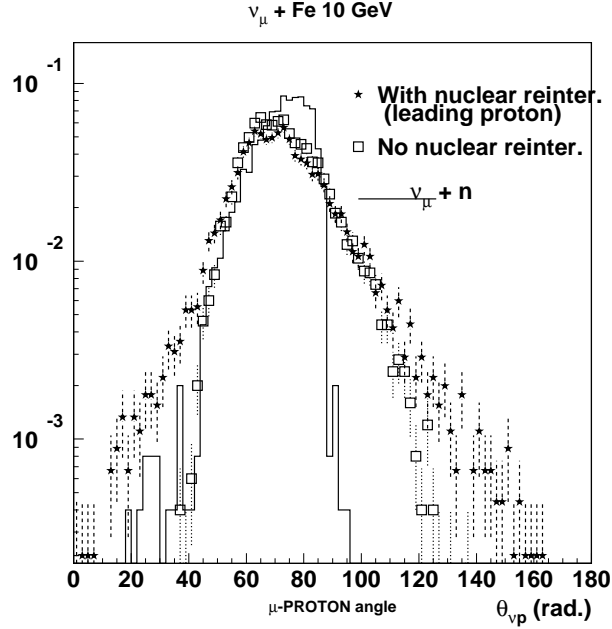,width=9cm}}
\caption {Distribution of the angle between the outgoing proton and the 
neutrino direction, in $\nu_{\mu}$+Fe 
interaction, $E_{\nu_{\mu}}$ = 10 GeV)}
\label{f:fig4}
\end{figure}
Also in naive models we expect the
appearance of non zero missing $P_{\perp}$ from Fermi motion. Nuclear 
re--interactions increase this effect ( Fig. 5 and 6 for C and Fe target
respectively).
\begin{figure}[htb]
\centerline{\epsfig{file=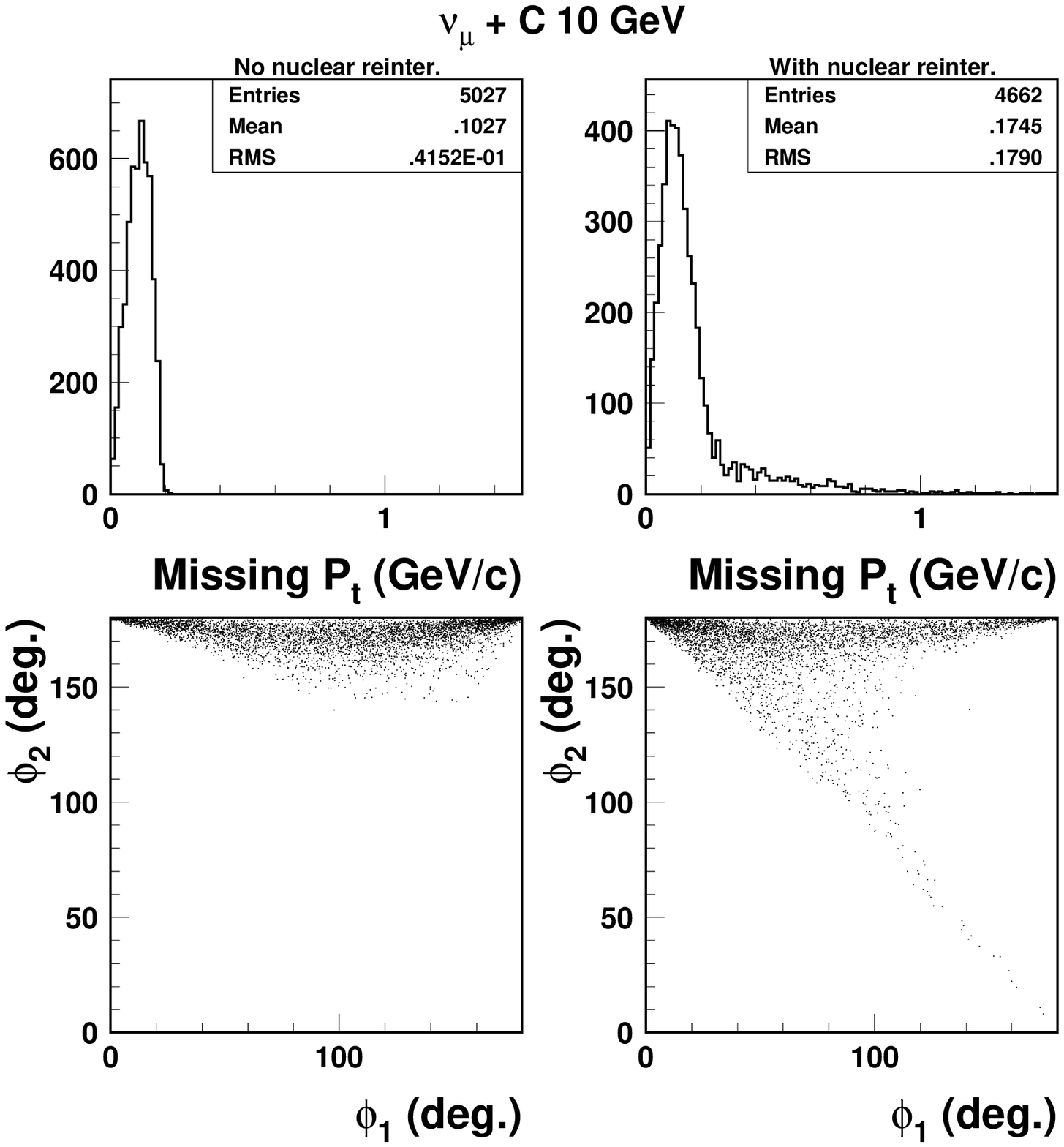,width=10cm}}
\caption {$\nu_{\mu}$+C interaction($E_{\nu_{\mu}}$=10 GeV).
Missing $P_{\perp}$ distribution (top) without (left) and with
nuclear re--interaction (right). Scatter plot of 
$\phi_{1}$ and $\phi_{2}$ angles (bottom) without (left) and with (right)
nuclear re--interaction.}                                                                                        
\label{f:fig5}
\end{figure}
There, we show in the top part the change in 
missing $P_{\perp}$ distribution
from the case without intranuclear cascade (left) to the case including
re--interactions (right). A tail in the last case is evident extending at
large value of $P_{\perp}$; this is a dangerous effect for the experiments 
aiming to tag the $\nu_{\tau}$ using kinematic cuts ``a la Nomad''\cite{nomad}.

\begin{figure}[htb]
\centerline{\epsfig{file=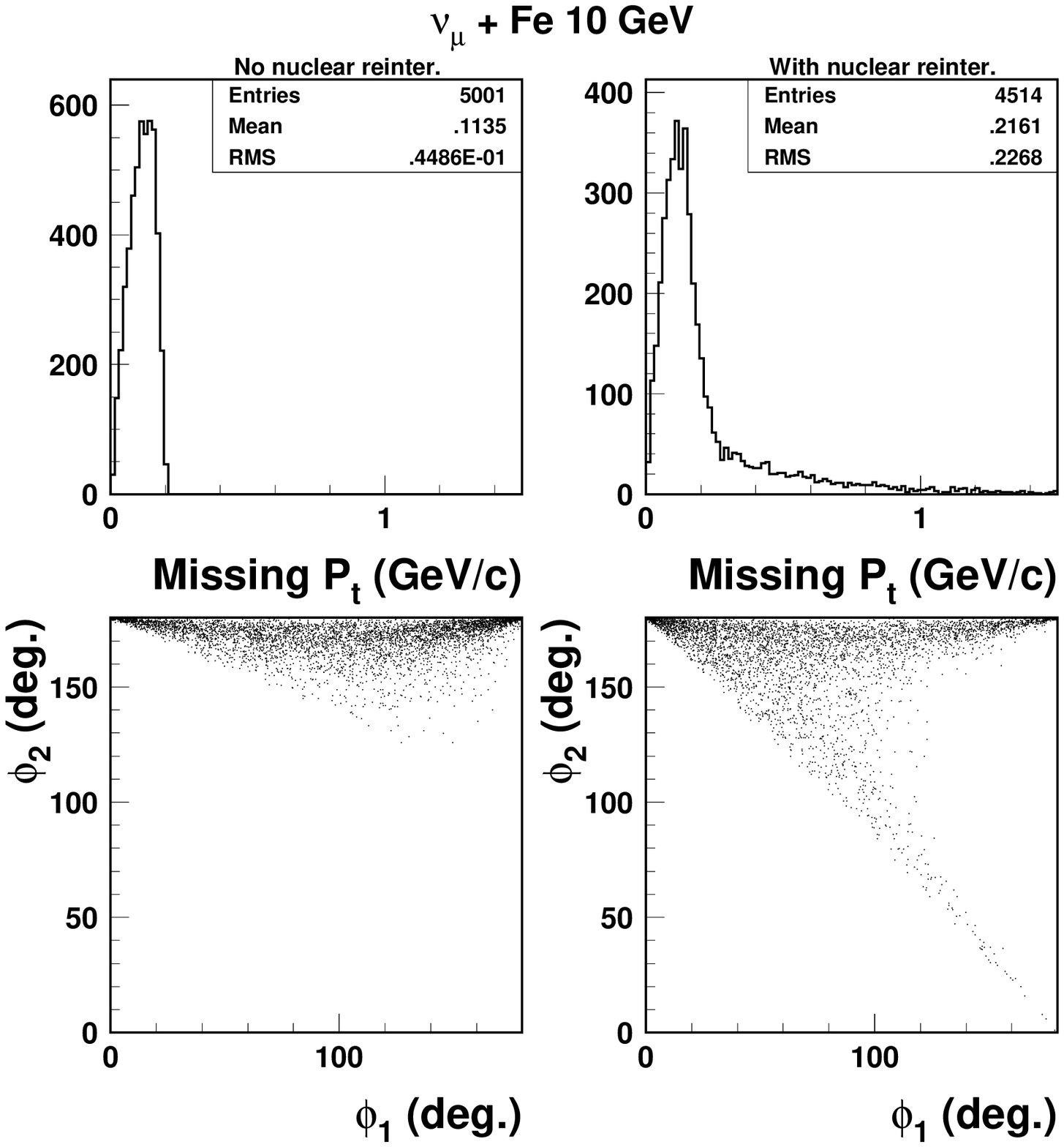,width=10cm}}
\caption {$\nu_{\mu}$+Fe interaction($E_{\nu_{\mu}}$=10 GeV. 
Missing $P_{\perp}$ distribution(top) without (left) and with
nuclear re--interaction (right). Scatter plot of 
$\phi_{1}$ and $\phi_{2}$ angles(bottom) without(left) and with (right)
nuclear re--interaction.}                                                                                        
\label{f:fig6}
\end{figure}       
This is clear in the bottom part of Fig. 5 and 6, where we show the 
scatter plot of  
the angle between the proton transverse momentum $P^{p}_{\perp}$ and the 
muon $P^{\mu}_{\perp}$($\phi_{2}$), versus the angle between $P^{p}_{\perp}$ 
and the missing transverse momentum $P^{miss}_{\perp}$ ($\phi_{1}$).
For completeness we show in Fig. 7
the case of $\nu_{\tau}$+n$\rightarrow$$\tau$+p,
where we considered the channel 
$\tau$$\rightarrow$$\mu$+$\nu$+$\bar{\nu}$
in Fe target. 
Due to
nuclear effects, we observe the presence of events in which $\nu_{\mu}$ 
interactions give small $\phi_{1}$ angles, close to the region expected
for $\nu_{\tau}$ candidates.
\begin{figure}[htb]
\centerline{\epsfig{file=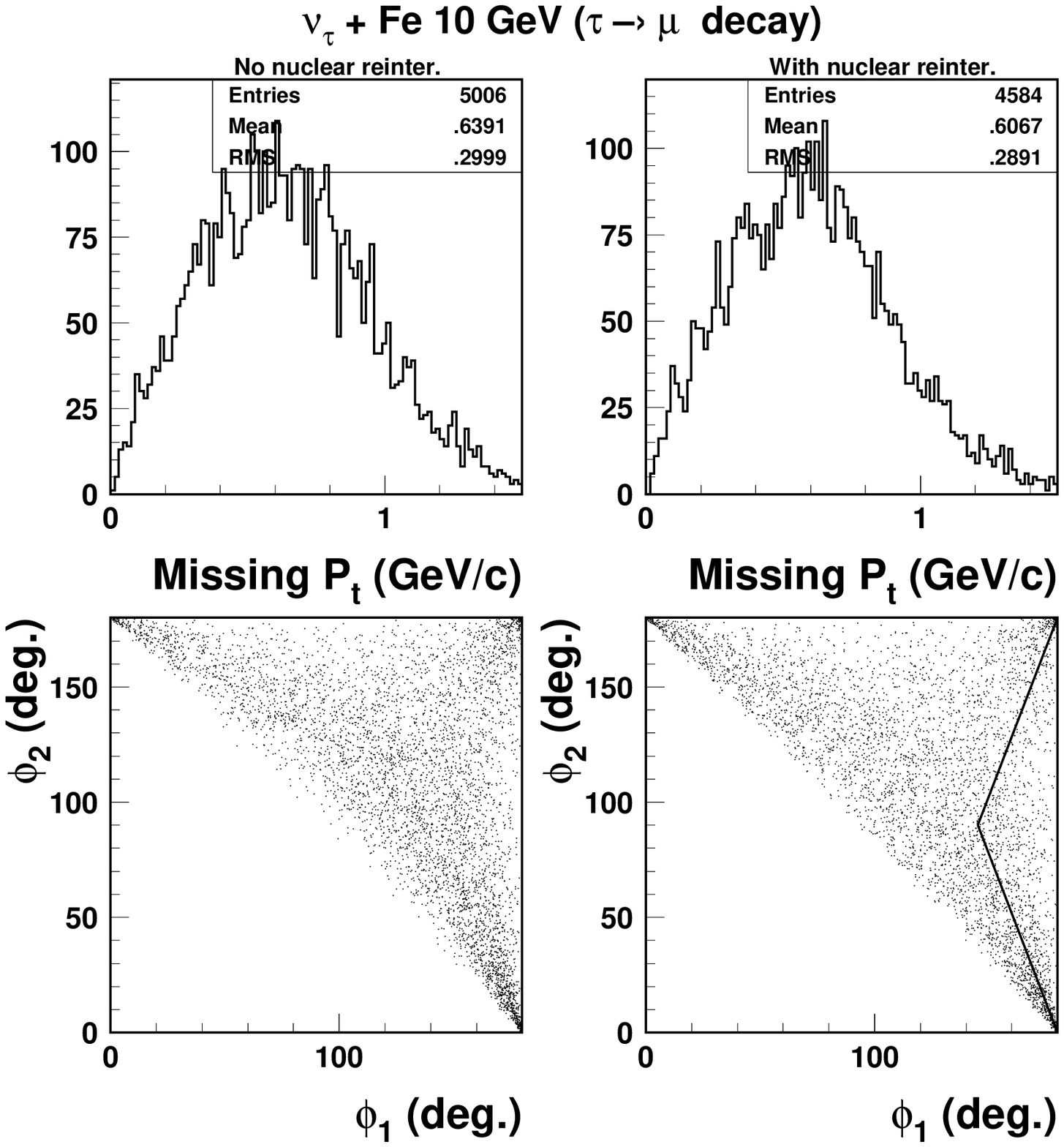,width=10cm}}
\caption {$\nu_{\tau}$+C interaction($E_{\nu_{\mu}}$=10 GeV. 
Missing $P_{\perp}$ distribution(top) without (left) and with
nuclear reinteraction(right). Scatter plot of 
$\phi_{1}$ and $\phi_{2}$ angles(bottom) without(left) and with (right)
nuclear reinteraction. A possible selection area to identify $\tau$ candidates
is the one to the right of the triangle drawn in the last plot.}  
\label{f:fig7}
\end{figure}       
                   
The intranuclear cascade is relevant as far as the 
$\nu_{\mu}$ and $\nu_{e}$ are concerned, while in this last case
($\nu_{\tau}$,$\tau$$\rightarrow$$\mu$+$\nu$+$\bar{\nu}$) it is obscured
by the missing $P_{\perp}$ introduced by the $\tau$ decay. 
In any case, it seems that, despite the smearing introduced by nuclear 
re--interactions, a region still exists in the $\phi_2$--$\phi_1$ plane which
allows the identification of $\tau$ candidates. 
Just to consider an example, although only at particle level,
if we take , the
(non optimized) region on the right of the triangle drawn in the 
rightmost bottom plot of
Fig. 7, we find that the efficiency of detecting a $\tau$ candidate
is 864/4584, having rejected 100\% of 4514 $\nu_\mu$ interactions.
Of course, this is just an over simplification, since in practice 
all the other resolution effects coming from a realistic experimental
simulation
have to be considered.

\section {Delta resonance} Delta resonance process has been included in our
code, for both CC and NC interactions.
As a relevant example for experimental application, we have considered
5,000 CC interations of 10 GeV $\nu_{\mu}$ on Iron. 
All the channels $\Delta^{++}$$\rightarrow$p+$\pi^{+}$,
$\Delta^{+}$$\rightarrow$p+$\pi^{0}$ and $\Delta^{+}$$\rightarrow$n+$\pi^{+}$
have been considered, and the results are shown in Table 2.
\begin{table}[hbt] 
\begin{center}
\begin{tabular}{|l|c|c|c|c|c|} 
\hline &&&&&
\\ Process & $<N_{p}>$ &
$<N_{n}>$&$<N_{\pi^{\pm}}>$&$<N_{\pi^{0}}>$&  $<N_{\gamma}>$  \\
\hline &&&&&
\\ $\Delta^{++}$$\rightarrow$p$\pi^{+}$&  2.14 & 1.97 & 0.85 &
0.03    &3.68        \\ \hline $\Delta^{+}$$\rightarrow$p$\pi^{0}$& 
2.13   & 1.57 & 0.05 & 1.02  & 3.46      \\ \hline
$\Delta^{+}$$\rightarrow$n$\pi^{+}$& 1.64   & 2.65 & 0.91 & 0.04  & 3.39 
\\ \hline 
\end{tabular} 
\caption{Comparison of the result of  $\nu_{\mu}$
interaction  ( $E_{\nu_\mu}$ = 10 GeV ) on Iron for different resonant
processes} 
\end{center}             
\end{table} 
As in the case of QEL
interaction, we see the presence of extra particles in the final state: in
practice, for Fe target,
we never observe events preserving the final state obtainable without
nuclear
reinteractions. It might be intersting to notice how in a non negligible
fraction of cases the charged pion can be absorbed inside the nucleus.
This complicates the separation of quasi elastic interactions from 
resonance excitation: these two classes must be always considered together.
Also,
the effect of the nuclear rescattering manifests in an additional tail
of event missing momentum.

\section{Conclusions} 
We have shown how the nuclear re--interaction  in quasi
elastic neutrino events can play an important role in the experimental neutrino
detection. The nuclear models contained in the DPMJET code seems to be adequate
for the description of the phenomenology of neutrino interaction in view of
design of future detectors for Long Base Line experiments. Of corse this is
just a preliminary stage of the work. Further  improvements will allow the
inclusion of  
scattering. Nuclear effect introduces also changes in the total
neutrino--nucleon cross section, as discussed in \cite{Sartogo}.
\vskip 2cm
\bibliographystyle{zpc} 
\bibliography{neutrino} 
\end{document}